\documentclass[useAMS,usenatbib]{mn2e}
\usepackage{graphicx}
\usepackage{epstopdf}

\title[A search for HeH$^+$ and CH]{A search for HeH$^+$ and CH in a high-redshift QSO}

\author[I. Zinchenko, V. Dubrovich \& C. Henkel]{I. Zinchenko$^{1,2}$\thanks{E-mail:
zin@appl.sci-nnov.ru (IAP)},
V. Dubrovich$^{3}$
and C. Henkel$^{4}$\\
$^1$Institute of Applied Physics of the Russian Academy of Sciences, Ulyanova 46,
603950 Nizhny Novgorod, Russia \\
$^2$Nizhny Novgorod University, Gagarin av. 23,
603950 Nizhny Novgorod, Russia \\
$^3$SPb Branch of SAO RAS, 196140, Pulkovskoe Shosse 65, St. Petersburg, Russia\\
$^4$Max Planck Institute for Radio Astronomy, Auf dem H\"{u}gel 69, P.O. Box 20 24,
53010 Bonn, Germany
}

\begin{document}

\date{ }

\pagerange{\pageref{firstpage}--\pageref{lastpage}} \pubyear{2011}

\maketitle

\label{firstpage}

\begin{abstract}
{We performed a search for the HeH$^+$ $J=1-0$ line ($\nu_{\rm rest}$ = 2010.183873\,GHz) and simultaneously for the CH $ ^2\Pi_{3/2}(F_2)J=3/2 -\, {^2\Pi}_{1/2}(F_2)J=1/2 $ lines ($\nu_{\rm rest}\approx$ 2006.8 and 2010.8\,GHz) toward one of the highest-redshift quasars known, SDSS J114816.64+525150.3 ($z= 6.4189$). No clearly visible line was detected after obtaining an rms noise level of $\sim 0.4 $~mK ($ \sim 3 $~mJy) in 16~MHz (18~km\,s$^{-1}$) channels. At a level of $2.9\sigma$, however, there is a tentative emission feature shifted by about 100~km\,s$^{-1}$ from the expected frequency of the HeH$^+$ line. This shift is well within the width of the line profiles for CO and C$^+$. The putative feature is about four times narrower than the previously detected CO and C$^+$ lines. The difference in velocities as well as in the line widths could be explained by quite different conditions required for formation and excitation of HeH$^+$ with respect to CO and C$^+$. The HeH$^+$ emission, if real, could probably arise in the dense ionized gas of this QSO. The velocity integrated flux in this tentative feature is $0.62\pm 0.21$~Jy$\cdot$km\,s$^{-1}$ which corresponds to a total luminosity of $ L(\mathrm{HeH^+})\approx 7.1\times 10^8 $~L$_{\sun}$. As long as there is no independent confirmation, these values should be considered rather as upper limits.}
\end{abstract}


\section{Introduction}
In the early universe, astrochemistry 
is still almost unexplored. Observations of dwarf galaxies with low
metallicities and spectral scans of ultraluminous infrared galaxies 
(ULIRGs) in the local universe are a great help, not 
only in getting a better understanding of these interesting classes
of targets, but also to improve our knowledge on molecular cloud properties 
and star formation under exotic conditions resembling to a certain extent those in the distant past. However, in the end -- in spite 
of all difficulties related to their large distances -- nothing can
replace direct detailed studies of the high-$z$ sources themselves.

While there have been some attempts to search for molecules other 
than CO at $z > 2$ (e.g. HCN, HNC, CN; \citealt{Riechers07,Wagg05,Weiss07,Guelin07}), the number of detected species is still fairly
small. Right now we only see the tip of the iceberg and a broader 
set of observational data is mandatory. One of the most promising 
molecules for studies of the early universe is HeH$^+$ \citep{Dubrovich95,Dubrovich97}. Hydrogen and helium are the main components of matter in the universe. Available chemical models predict HeH$^+$ to be one of the most abundant molecules in the post-recombination era \citep[e.g.][]{Bovino11}. 

Several authors analysed HeH$^+$ formation and excitation in astrophysical plasmas and in various nebulae \citep[e.g.][and references therein]{Roberge82,Cecchi93}. They found that the HeH$^+$ abundance can be particularly high in vicinity of strong extreme UV and X-ray emitting sources, sufficient to produce an observable line emission.

Despite of the expectations this molecule has not been detected in space yet. With the ISO LWS instrument \citet{Liu97} detected an emission feature at the frequency of the HeH$^+$ $J=1-0$ transition at 149.18~$\mu$m in the planetary nebula NGC7027. However, their analysis shows that this feature is most probably attributed to the CH $ ^2\Pi_{3/2}(F_2)J=3/2 -\, {^2\Pi}_{1/2}(F_2)J=1/2 $ fundamental pure rotational lines at nearby frequencies (the ISO LWS spectral resolution was insufficient to separate these lines). 

Nevertheless, we may expect a rather high abundance of HeH$^+$ in dense clouds near distant quasars where its cosmological abundance can be drastically enhanced by local energetic processes. Some other simple molecules (e.g. CH) may also be sufficiently abundant.

We report here results of a search for HeH$^+$ and CH in the spectrum of one of the highest-redshift quasars, SDSS J114816.64+525150.3 ($z= 6.4189$). A rather strong emission in the redshifted $^2P_{3/2} - {^2}P_{1/2}$ fine-structure line of C$^+$ at 157.74 $\mu$m was detected there using the IRAM 30-m telescope \citep{Maiolino05}, in addition to CO emission detected earlier \citep{Walter03,Walter04,Bertoldi03a}. Here we report on a search for the HeH$^+$\,(1--0) line ($\nu_{\rm rest}$ = 2010.183873\,GHz, \citealt{Matsushima97}) and simultaneously for the CH $ ^2\Pi_{3/2}(F_2)J=3/2 -\, {^2\Pi}_{1/2}(F_2)J=1/2 $ lines ($\nu_{\rm rest}\approx$ 2006.8 and 2010.8\,GHz, \citealt{Pickett98}) in this object. Due to the redshift these lines fall into the frequency interval from 270.5~GHz to 271~GHz which can be readily observed at the 30-m IRAM telescope.

\section{Observations}
The observations were performed in January 2011 in a pool observing session at the IRAM 30m radio telescope in the dual beam switching mode using wobbler switching with a beam throw of $\pm 60\arcsec$ towards the nominal source position of $\rmn{RA}(2000)=11^{\rmn{h}} 48^{\rmn{m}} 16\fs6$, $\rmn{Dec.}~(2000)=52\degr 51\arcmin 50\arcsec$. The receiver at both polarizations was tuned to the red-shifted HeH$^+$ frequency, equal to 270.954437~GHz. The system temperature varied from $\sim 180$~K to $\sim 400$~K. The total integration time was about 20 hours. The antenna HPBW at this frequency is about 9~arcsec. Pointing was checked approximately every 1.5 hours and was found to be stable within $\sim 4$ arcsec. The results are presented on an antenna temperature scale, $T_{\rm A}^*$, corrected for aperture blocking and for atmospheric attenuation.

We used two spectral backends in parallel: the WILMA autocorrelator with a 2~MHz channel width and the filter bank with a 4~MHz channel width. Both backends had 4~GHz total bandwidth. 

The data reduction was performed with the GILDAS software package. The data set consists of about 250 individual spectral scans. Each scan was checked visually; several apparently corrupted scans were omitted. With the WILMA backend many scans show spurious spikes in a few individual channels. However these channels are located well outside the expected frequency range for our lines and no special measures were made to exclude these spikes. On the other hand some spectra obtained with the filter bank suffer from a strong baseline curvature on rather large scales which can affect final results.

\section{Results}

\begin{figure}
  \includegraphics[angle=-90,width=\columnwidth]{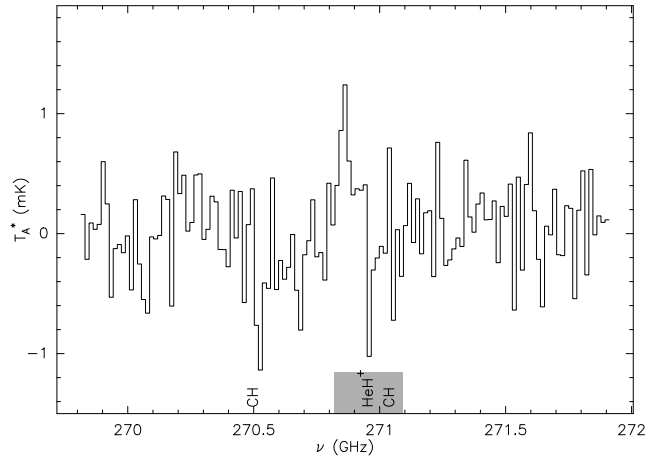}\\
  \caption{An average spectrum obtained with the WILMA autocorrelator smoothed to a 16~MHz resolution after subtraction of a zero-order baseline. The expected frequencies of the HeH$^+$ and CH lines are indicated. The shaded area shows the frequency range approximately corresponding to the width of the CO and C$^+$ lines at the half intensity level (300~km\,s$^{-1}$).}\label{fig:w-aver}
\end{figure}

An average spectrum obtained with the WILMA autocorrelator smoothed to a 16~MHz (18~km\,s$^{-1}$) resolution after subtraction of a zero-order baseline is shown in Fig.~\ref{fig:w-aver}. The expected frequencies of the HeH$^+$ and CH lines are indicated. The shaded area shows the frequency range approximately corresponding to the width of the CO and C$^+$ lines at the half intensity level (300~km\,s$^{-1}$).

According to the common detection criterion ($ S/N > 3 $) there is no detected line in this spectrum. The rms noise in this 16~MHz resolution spectrum is $\sim 0.4 $~mK ($ \sim 3 $~mJy using a conversion factor of 8.4~Jy/K).

At the same time one can see a possible emission feature in our spectrum close to the expected HeH$^+$ frequency. A formal Gaussian fit gives for the integrated intensity $67\pm 23$~mK$\cdot$MHz or $0.62\pm 0.21$~Jy$\cdot$km\,s$^{-1}$, i.e. the signal to noise ratio is about 2.9. The line width is $61\pm 32$~MHz or $68\pm 36$~km\,s$^{-1}$. The feature is shifted by $103\pm 10$~km\,s$^{-1}$ from the expected frequency of the HeH$^+$ line. As mentioned above this shift is well within the width of the line profiles for CO and C$^+$.

\begin{figure}
  \includegraphics[angle=-90,width=\columnwidth]{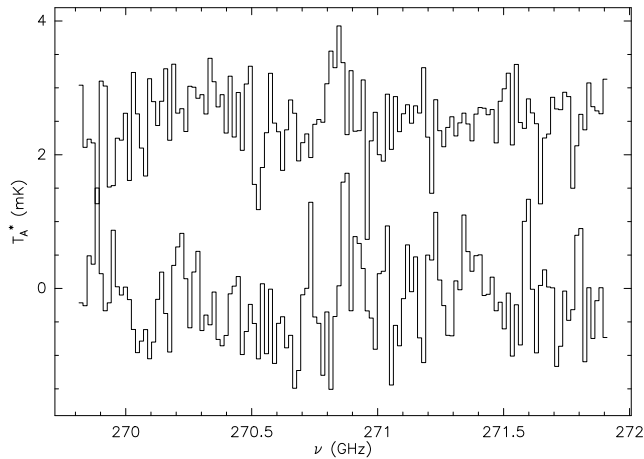}\\
  \caption{Average spectra of the two halves of the data set obtained with the WILMA autocorrelator smoothed to a 16~MHz resolution. One of the spectra is shifted along the vertical axis for clarity.}\label{fig:2halves}
\end{figure}

\begin{figure}
  \includegraphics[angle=-90,width=\columnwidth]{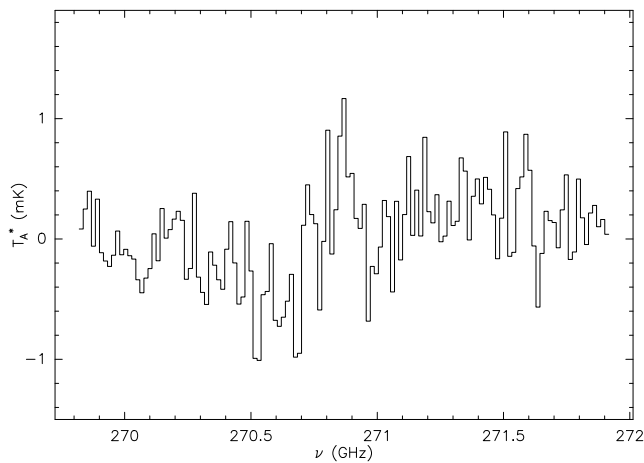}\\
  \caption{The average spectrum obtained with the filter bank smoothed to a 16~MHz resolution after subtraction of a zero-order baseline.}\label{fig:4m}
\end{figure}

In order to check further the significance of this result we split the data set into two approximately equal parts and analysed them separately. The average spectra for these parts are shown in Fig.~\ref{fig:2halves}. Both spectra show hints for the emission features in the same frequency range. The same feature is seen in the spectrum obtained with the filter bank (Fig.~\ref{fig:4m}) although the baseline here is worse than in the WILMA spectrum. We conclude that we have obtained a very tentative detection of an emission line which can be attributed to HeH$^+$, while there is no sign for CH emission or absorption. 

\section{Discussion}

What could be the reason for the differences in velocity and width between this tentative HeH$^+$ line and the detected lines of CO and C$^+$?
As shown by \citet{Roberge82,Cecchi93}  the HeH$^+$ emission should arise mostly from edges of dense ionized regions. According to these works HeH$^+$ can be effectively excited by electron impacts. The required electron density is $ n_\mathrm{e} \ga 10^5-10^6 $~cm$^{-3}$. Collisional excitation of HeH$^+$ by neutrals requires very high densities and seems to be unlikely. Therefore, our tentatively detected line could be produced in dense ionized regions in this QSO. It would then not be surprising that its central velocity and velocity dispersion are different from those of the neutral molecular gas traced by CO and C$^+$. For example, high ionization UV lines in this object have a velocity offset of $ \sim 2000 $~km\,s$^{-1}$ \citep{Walter03}. It is worth mentioning also that presumably strong UV and X-ray emissions of this very luminous QSO, which probably harbours a $ 3\times 10^9 $~M$_{\sun}$ black hole in the center \citep*{Willott03}, create favourable conditions for HeH$^+$ formation.

The total luminosity in the line, if real, can be estimated as \citep{Solomon97}
\begin{equation}
L=1.04\times 10^{-3} D_{\mathrm{L}}^2 \nu_{\mathrm{rest}}(1+z)^{-1} \int S\,dv ,
\end{equation}
where $ L $ is the line luminosity in L$_{\sun}$, $ \nu_{\mathrm{rest}} $ is the line rest frequency in GHz, $ D_{\mathrm{L}} $ is the luminosity distance in Mpc and $ \int S\,dv $ is the velocity integrated flux density in Jy\,km\,s$^{-1}$. Using the Cosmology Calculator \citep{Wright06} with the ``standard'' cosmological parameters ($H_0=71$~km\,s$^{-1}$Mpc$^{-1}$, $\Omega_\mathrm{M}=0.27$, $\Omega_\mathrm{\Lambda}=0.73$) we obtain $ D_{\mathrm{L}}=63776 $~Mpc. 
Then, $ L(\mathrm{HeH^+})\approx 7.1\times 10^8 $~L$_{\sun}$. Until a reliable confirmation the estimates of the line intensity and luminosity should be considered rather as upper limits.

\section{Conclusions}

Our data provide no clear detection of either HeH$^+$ or CH in a high-redshift ($z= 6.4189$) QSO SDSS J114816.64+525150.3 where a rather strong emission in several atomic and molecular lines has been detected earlier. However, there is a possible (at a level of $2.9\sigma$) emission feature which can be attributed to the HeH$^+$ $ J=1-0 $ line.  This feature is shifted by about 100~km\,s$^{-1}$ from the expected frequency of the HeH$^+$ line derived from the CO and C$^+$ line redshifts and is about 4 times narrower than these lines. This shift is well within the width of the line profiles for CO and C$^+$. The difference in velocities as well as in the line widths can be explained by quite different conditions required for formation and excitation of HeH$^+$ on the one hand and CO as well as C$^+$ on the other hand. The HeH$^+$ emission could probably arise in dense ionized gas in this QSO. The velocity integrated flux in this tentative line is $0.62\pm 0.21$~Jy$\cdot$km\,s$^{-1}$ which corresponds to the total luminosity in the line of $ L(\mathrm{HeH^+})\approx 7.1\times 10^8 $~L$_{\sun}$. Prior to a reliable confirmation, these values should be considered rather as upper limits.

\section*{Acknowledgements}
We are very grateful to the IRAM director, Pierre Cox for providing Director's Discretionary Time for this project and to the IRAM staff, especially to Manuel Gonzalez and Clemens Thum for their help and advices with the observations. We thank Sergey Levshakov for attracting our attention to this object.

The work was supported by
Russian Foundation for Basic Research and by the Russian Academy of Sciences.
The research has made use of the SIMBAD database,
operated by CDS, Strasbourg, France.

\bsp

\label{lastpage}

\end{document}